\documentclass[aps,pra,onecolumn,nofootinbib,superscriptaddress]{revtex4}
\usepackage{amsmath}
\usepackage{amsfonts}
\usepackage{amssymb}
\usepackage{graphicx}
\usepackage{comment}
\usepackage{color}

\begin{document}

\title{Mutually Unbiased Unitary Bases}

\author{Jesni Shamsul Shaari}
\affiliation{Faculty of Science, International Islamic University Malaysia (IIUM),
Jalan Sultan Ahmad Shah, Bandar Indera Mahkota, 25200 Kuantan, Pahang, Malaysia}
\author{Rinie N. M. Nasir}
\affiliation{Faculty of Science, International Islamic University Malaysia (IIUM),
Jalan Sultan Ahmad Shah, Bandar Indera Mahkota, 25200 Kuantan, Pahang, Malaysia}
\author{Stefano Mancini}
\affiliation{School of Science \& Technology, University of Camerino, I-62032 Camerino, Italy}
\affiliation{ INFN Sezione di Perugia, I-06123 Perugia, Italy}

\date{\today}

\begin{abstract}
 We consider the notion of unitary transformations forming bases for subspaces of $M(d,\mathbb{C})$ such that the square of Hilbert-Schmidt inner product of matrices from the differing bases is a constant. Moving from the qubit case, weconstruct the maximal number of such bases for the 4 and 2 dimensional subspaces while proving the nonexistence of such a construction for the 3 dimensional case.  Extending this to higher dimensions, we commit to such a construct for the case of qutrits and provide evidence for the existence of such unitaries for prime dimensional quantum systems. Focusing on the qubit case, we show that the average fidelity for estimating any of such a transformation is equal to the case for estimating a completely unknown unitary from $\text{SU}(2)$. This is then followed by a quick application for such unitaries in a quantum cryptographic setup.
\end{abstract}

%\pacs{03.65.Aa, 03.67.-a}
%\keywords{}

\maketitle
\section{Introduction}
\noindent The idea of indistinguishability of quantum states and the inability to clone an arbitrary state has brought with it the acceptance of physical limits on how much one can know of a quantum system. In the context of mutually unbiased bases (MUB) - for a review on the subject of MUB see e.g. ref.\cite{durt}, considering an unknown quantum state selected from sets of MUBs; a measurement of an observable described by projectors onto states in a basis mutually unbiased with respect to the prepared one  produces completely random results. This would later come to be the key ingredient in the well known BB84 quantum key distribution (QKD) protocol \cite{BB84} as well as most (if not all) prepare and measure QKD schemes where eavesdropping of a transmitted message would statistically induce errors for which below a certain threshold, legitimate parties may distill a secret key nonetheless.

While the optimal estimation of a quantum state deals with the maximal information extraction of the system's state, the issue of estimating an unknown physical transformation, described by a unitary operator on a state is another matter. This was studied in \cite{acin} where it is a more challenging task compared to state estimation as it requires not only the use of optimal measurements of the resulting state post transformation but also an optimal state prior to the transformation to begin with. An immediate use of such estimation, namely the alignment of reference frames using quantum spins was noted in \cite{chiri2004} as being equivalent to estimating unknown rotations of $\text{SU}(2)$.  

This paradigm has also found its way in the field of quantum cryptography where protocols making use of a bi-directional quantum channel and unitary transformations for encoding purposes like that of \cite{PP,LM, TCS} rely on the inability of an eavesdropper (commonly referred to as Eve), to ascertain the transformations perfectly without inducing errors. However, this inhibition for Eve does not reflect an intrinsic inability with regards to the transformation themselves, rather is chiefly due to the inability of estimating the encoded states traveling between the legitimate parties (commonly referred to as Alice and Bob). It was in \cite{chiri2} that the notion of two-way QKD based on indistinguishable unitaries was first mooted where Alice's transformation for encoding purposes is selected from two orthogonal bases of unitary transformations; one comprising the identity and Pauli matrices, the other would be those unitaries  multiplied by a rotation by an angle $2\pi/3$ about the axis $(1,1,1)/\sqrt{3}$ of $\mathbb{R}^3$. In \cite{bisio}, these bases were referred to as mutually unbiased
bases of orthogonal qubit unitaries. A two-way QKD study with non-entangled qubits using `nonorthogonal' unitaries was later reported in \cite{jss}, providing for a higher security threshold compared to its conventional predecessors for selected attack strategies. 

Such a notion of ``nonorthogonal" unitaries  is intimately linked to unitary 2-designs \cite{scott, roy} which is a set of matrices such that the average over it reproduces the average over the entire unitary group. The notion of `mutually unbiased unitary-operator bases' (MUUB) was put forward in \cite{scott} for unitary operators acting on a $d$ dimensional Hilbert space as the set of $d^2 - 1$ unitary operator bases with the property that each pair is \textit{mutually unbiased}. That is two sets of operator bases (each of cardinality $d^2$), say $\mathcal{T}_0$ and $\mathcal{T}_1$, are mutually unbiased if
\begin{eqnarray}\label{sd}
\left|\text{Tr}(T_0^{(i)\dagger}T_1^{(j)})\right |^2=1~~,~~\forall T_0^{(i)}\in\mathcal{T}_0, T_1^{(j)}\in\mathcal{T}_1, 
\end{eqnarray}
with $i,j=1,\ldots, d^2$ and $\text{Tr}(A^\dagger B)$ is the Hilbert-Schmidt inner product of $A$ and $B$.
It was used in the construction of an unitary 2-design, a study done in the context of process tomography of unital quantum channels. Ref.\cite{scott} also showed that for $d=2,3,5,7$ and $11$, a complete set, i.e. $d^2-1$ MUUB, can be found. The idea of such unbiasedness for a unitary basis is arguably deeper than afforded by the definition of 2-designs and in our work here, we treat the notion of such unbiased unitary bases in its own right.

Motivated by the equiprobable transition between states in one basis to another in the case of MUB, we consider an analogous idea of equiprobable guesses of unitaries towards a generalized notion of unbiased unitary operator bases (% referring to them as generalized mutually unbiased unitary bases, though  for short 
though we retain the abbreviation, MUUB) and provide a systematic study of their properties. The generalization here is made in the sense of neither restricting the value of the Hilbert-Schmidt norm of eq.(\ref{sd}) to unity nor the cardinality of the operator basis. We proceed in Section II, with the characterization on the maximal number of MUUB on a two dimensional Hilbert space, $\mathcal{H}_2$, namely qubit. We then show how this is immediately connected to maximally entangled states forming MUBs (Section III), explicitly so for the relevant subspaces of the Hilbert space $\mathcal{H}_2\otimes \mathcal{H}_2$. We then provide in the subsequent section, an explicit construction for the case of MUUB on a 3-dimensional Hilbert space and follow up with a simple proof of the existence of MUUBs on prime numbered dimensional quantum systems. Focusing on the MUUBs for $\mathcal{H}_2$, we calculate the fidelities in distinguishing between the unitaries. We show how this is equal to the discrimination of the maximally entangled states from the varying MUBs as well as the estimation of a completely unknown unitary from $\text{SU}(2)$  (Section IV). Before concluding, we discuss the use of MUUB in a QKD setup.

%%%%%%%%%%%%%%%%%%%%%%%%%%%%%%%%%%%%%%%%%%%%%%%%%%%%%%%%%%%%%%%%%%%%%%%%%%%%%%%%%%%%

\section{{ MUUB} for Qubits}

MUB refers to orthonormal bases of a Hilbert space such that the transition from any one state in one basis to any state in the other basis are equiprobable \cite{durt}. In other words, if one is asked to guess an unknown state, $|u\rangle$ prepared in one basis of say a $d$ dimensional Hilbert space, $\mathcal{H}_d$,  and we let the states $|r\rangle$ and $|s\rangle$ be `guesses' (resulting from a measurement) from a mutually unbiased basis, then 
\begin{eqnarray}
|\langle u|r\rangle|^2=|\langle u|s\rangle|^2=1/d.
\end{eqnarray}   
In estimating an unknown unitary $U\in \text{SU}(d)$ with a guess  $U_r$, Ref.\cite{acin} considers how closely $U_r$ resembles $U$ by determining the behaviour of the unitaries, averaged over all states $|\alpha\rangle\in \mathcal{H}_d$ given by,  
\begin{eqnarray}
\left|\int\langle\alpha|U_r^\dagger U|\alpha\rangle d\alpha\right|^2=\dfrac{1}{d^2}|\text{Tr}(UU_r^\dagger)|^2.
\end{eqnarray}
Thus, beginning with a unitary $U$, let $U_{r}$ and $U_{s}$ be guesses for $U$, we could say that the guesses are equiprobable if $|\text{Tr}(UU_{r}^\dagger)|^2=|\text{Tr}(UU_{s}^\dagger)|^2$.\\
\newline
\noindent{\bf Definition 1}\\
Consider two distinct orthogonal basis, $\mathcal{A}_0,\mathcal{A}_1$ comprising of unitary transformations for some subspace of the vector space $M(d,\mathbb{C})$%such that matrices in each set are pairwise orthogonal; i.e. $\text{Tr}(b_k^{i*}b^j_k)=0,~\forall b^i_k, b^j_k\in\mathcal{B}_k$
. $\mathcal{A}_0$ and $\mathcal{A}_1$ are sets of MUUB provided  
\begin{eqnarray}
\left|\text{Tr}(A_0^{(i)\dagger}A_1^{(j)})\right |^2=C~~,~~\forall A_0^{(i)}\in\mathcal{A}_0, A_1^{(j)}\in\mathcal{A}_1, 
\end{eqnarray}
for $ i,j,=1,\ldots, n$ and some constant $C\neq 0$. \\

%In the following, we shall consider only two types of operators which could be MUO, namely Hermitian (which corresponds to observables) and unitary operations (corresponding to physical evolutions of states). 
With the number $n$ as the cardinality of the basis sets which reflect the dimensionality of the subspace,  in the following  subsections, we limit ourselves to unitaries acting on states in $\mathcal{H}_2$, i.e. $d=2$ and provide explicit constructions of MUUB for $n=4$ and $2$. Furthermore we note that no MUUB exist for $n=3$. This is simply due to the fact that any subspace spanned by three orthogonal $2\times 2$ matrix necessitates the uniqueness of such a basis. \\

We should like to note that, differently from \cite{scott}, the definition given above is motivated by the idea of equiprobable guesses of a unitary and is more general in the sense that the definition in \cite{scott} requires the Hilbert-Schmidt inner product of the matrices to equal $1$, restricted to the subspace of dimension $d^2$; ours is a constant $C$ without specifying the dimensionality of the subspace. We can see later, at least through the explicit construction of MUUB acting on $\mathcal{H}_2$, how our definition reduces to \cite{scott} in the case of $n=4$ while the value of $C$ is different in the case of $n=2$. We further note that %while \cite{scott} mentioned that MUUs is known to exist for dimensions $2,3,5,7$ and $11$ (for the largest subspace with dimension $d^2$), 
our constructions for MUUB is straightforwardly derived from our definition and should possibly provide for simpler insights to higher dimensions.

%%%%%%%%%%%%%%%%%%%%%%%%%%%%%%%%%%%%%%%%%%%%%%%%%%%%%%%%%%%%%%%%%%%%%%%%%%%%%%%%%%%%

\subsection{MUUB for $n=4$}

\noindent We will first need to ensure that our construction here, would provide for a result which holds in general. To this end, we define the equivalence between bases for possibly distinct subspaces.\\  

\noindent{\bf Definition 2}\\
Consider two distinct orthogonal basis, $\mathcal{X}=\{X_1,X_2,...,X_n\}$ and $\mathcal{Y}=\{Y_1,Y_2,...,Y_n\}$ for some $n$ spanning possibly distinct subspaces $M(\mathcal{X})$ and $M(\mathcal{Y})$ respectively. $\mathcal{X}$ is equivalent to $\mathcal{Y}$, $\mathcal{X}\equiv \mathcal{Y}$, provided that $\zeta_j Y_j=UX_j, \forall j=1,2,...,n$ for some unitary $U$ and $|\zeta_j|=1$.\\ 

\noindent{\bf Proposition 1}\\
Given the above definition for $\mathcal{X}\equiv \mathcal{Y}$ for some $n$, $\mathcal{X}$ is unbiased with regards to a set $\mathcal{Z}=\{Z_1,Z_2,...,Z_n\}\subset M(\mathcal{X})$ if and only if $\mathcal{Y}$ is unbiased with regards to a set  $U\mathcal{Z}=\{UZ_1,UZ_2,...,UZ_n\}\subset M(\mathcal{Y})$.\\ 

\noindent The above proposition is easily shown to be true by $|\text{Tr}(X_j^\dagger Z_k)|^2=|\text{Tr}((UX_j)^\dagger UZ_k)|^2=|\text{Tr}((\zeta_j Y_j)^\dagger UZ_k)|^2=|\text{Tr}((Y_j)^\dagger UZ_k)|^2$. 

The case for $n=4$ here has $M(\mathcal{X})=M(\mathcal{Y})$ and is trivially the case of changing basis in $M(2,\mathbb{C})$. Thus, one could begin with a basis containing arbitrary but pairwise orthogonal unitaries for $M(2,\mathbb{C})$, and do a basis change by multiplying each element of the basis with some unitary operator (and ignoring possible overall phase factors) resulting in the basis $\mathcal{B}_0=\{\mathbb{I}_2,\sigma_1,\sigma_2,\sigma_3\}$ where $\mathbb{I}_2$ is the identity operator and $\sigma_1,\sigma_2,\sigma_3$ are the Pauli matrices given by
\begin{eqnarray}
\sigma_1=\begin{pmatrix}
  0 & 1\\
  1 & 0
   \end{pmatrix}~,~
\sigma_2=\begin{pmatrix}
  0 & -i\\
  i & 0
   \end{pmatrix}~,~
\sigma_3=\begin{pmatrix}
  1 & 0\\
  0 & -1
   \end{pmatrix}.
\end{eqnarray}
We can therefore consider the basis $\mathcal{B}_0$ with no loss of generality.
We next define a unitary element $U$ as a linear combination of elements of $\mathcal{B}_0$,
\begin{eqnarray}\label{U}
U=p_0\mathbb{I}_2+\sum_i^3p_i\sigma_i.
\end{eqnarray} 
Along with $\left|\text{Tr}(\sigma_jU)\right |^2=4|p_j|^2, \forall j$ and $U^\dagger U=\mathbb{I}_2$, we have 
$|p_j|^2=1/4$.
On the other hand, a generic $2\times 2$ unitary matrix, $U$, may be written as 
\begin{eqnarray}\label{G}
U=e^{i\alpha}\begin{pmatrix}
  a & b\\
  -b^* & a^*
 \end{pmatrix},
\end{eqnarray}
 with $|a|^2+|b|^2=1$.
By comparing eq.(\ref{U}) and eq.(\ref{G}) with regards to element $e^{i\alpha}a$ and $e^{i\alpha}a^*$, we can write 
\begin{eqnarray}
e^{i\alpha}a=p_0+p_3\Rightarrow e^{i\alpha}a^*=e^{2i\alpha}(p_0^*+p_3^*) ,\\
e^{i\alpha}a^*=p_0-p_3\Rightarrow e^{i\alpha}a=e^{2i\alpha}(p_0^*-p_3^*) .
\end{eqnarray}
The above equations give us 
\begin{eqnarray}
p_0^2=-p_3^2=e^{2i\alpha}/4.
%p_0=e^{2i\alpha}(p_0^*)\Rightarrow p_0^2=e^{2i\alpha}/4, \\
%p_3=-e^{2i\alpha}(p_3^*)\Rightarrow p_3^2=-e^{2i\alpha}/4 .
\end{eqnarray}
Similar treatment for elements $e^{i\alpha}b$ and $-e^{i\alpha}b^*$ would result in
\begin{eqnarray}
p_1^2=p_2^2=-e^{2i\alpha}/4.
%p_2^2=-e^{2i\alpha}/4 .
\end{eqnarray}
There can only be $2$ solutions for each $p_i$ in the above equations.%, corresponding to positive and negative solutions.
 Representing the solutions for $p_i$ to the above equations as a $4$-vector, $(p_0,p_1,p_2,p_3)$, one can have at most $16$ set of solutions for a given $\alpha$. As
\begin{eqnarray}\label{equi}
\left|\text{Tr}(U_a^{\dagger}(e^{i\beta}U_b))\right |^2=\left|\text{Tr}(U_a^{\dagger}(U_b))\right |^2,
\end{eqnarray}
we say $e^{i\beta}U_b$ is equivalent to $U_b$; %This implies that a matrix is mutually unbiased with respect to $e^{i\beta}U_b$ if and only if it is mutually unbiased to $U_b$ and 
$e^{i\beta}$ acts like a `global phase factor' and is inconsequential in the context of MUUB. 
Hence, the  set $(p_0,p_1,p_2,p_3)$ and $(-p_0,-p_1,-p_2,-p_3)$ (the case where $\beta=\pi$) would provide for equivalent matrices, then the number of nonequivalent solutions would be at most $8$  for a given $\alpha$. As the factor $\alpha$ itself has no effect on the equivalence of matrices, we conclude that there can be at most $8$ matrices which are mutually unbiased with regards to all elements in $\mathcal{B}_0$.
The $8$ set of solutions allow us to explicitly construct the vectors for bases $\mathcal{B}_1$ and $\mathcal{B}_2$ as
\begin{eqnarray}
%\mathcal{B}_1=\left\{\dfrac{\mathbb{I}_2+\sum_{j=1}^3ic_j \sigma_j}{2}|~c_j\in\{\pm1\},\prod_{j=1}^3{c_j}=-1\right\},\\\nonumber
\mathcal{B}_k=\left\{\dfrac{\mathbb{I}_2+i\sum_{j=1}^3d_j \sigma_j}{2}~|~d_j\in\{\pm1\},\prod_{j=1}^3{d_j}=(-1)^k\right\}\\\nonumber
~~~~~~~~~~~~~~~~~~~~~~~~~~~~~~~~~~~~~~~~~
\end{eqnarray}
for $k=1,2$. It can easily be shown that both $\mathcal{B}_1$ and $\mathcal{B}_2$ span $M(2,\mathbb{C})$ and
\begin{eqnarray}
\left|\text{Tr}(U_a^{\dagger}U_b)\right |^2=1~~,~~\forall U_a\in \mathcal{B}_1, U_b\in \mathcal{B}_2.
\end{eqnarray}
Thus the number of MUUB for $n=4$ is $3$ (including $\mathcal{B}_0$). This maximal number is known to be achieved from \cite{scott}.
%%%%%%%%%%%%%%%%%%%%%%%%%%%%%%%%%%%%%%%%%%%%%%%%%%%%%%%%%%%%%%%%%%%%%

\subsection{MUUB for n=3}

\noindent As mentioned earlier, no MUUB can exist in the case of $n=3$. The proof is rather straightforward. %This is simply due to the fact that the orthogonal basis containing $3$ elements that spans a subspace is necessarily unique (up to a universal operation). While we do not explicitly show proof of this (an exercise in linear algebra), we provide a brief sketch of the proof. 

Beginning with an arbitrary basis $\mathcal{V}=\{V_1,V_2,V_3\}\subset M(2,\mathbb{C})$ for a $3$ dimensional subspace of $M(2,\mathbb{C})$, it can always be mapped to $\mathcal{S}=\{\mathbb{I}_2,\sigma_i,\sigma_j\}$, for $i \neq j$ where $V_1^\dagger V_1=\mathbb{I}_2$ and $V_1^\dagger V_2=\sigma_i$. If we require $\mathcal{V}$ to be  unbiased with respect to $\mathcal{S}$, $V_1$ must be an element of the span of $\mathcal{S}$ and $V_1=q_0\mathbb{I}_2+q_i\sigma_i+q_j\sigma_j$ with $|q_0|^2=|q_1|^2=|q_2|^2=1/3$. Thus $V_2=V_1\sigma_i=q_0\sigma_i+q_i\mathbb{I}_2+q_j\sigma_i\sigma_j$. However, $\sigma_i\sigma_j$ is not in the span of $\mathcal{S}$, hence $\mathcal{V}$ cannot be an MUUB with respect to $\mathcal{S}$ and we conclude no MUUB exist for $n=3$. %$V$ should necessarily map $\{\mathbb{I}_2,A^\dagger B,A^\dagger C\}$ to another basis. However, it can be shown that one element in the set cannot be described by a linear combination of the elements in $\{\mathbb{I}_2,A^\dagger B,A^\dagger C\}$.
%{\red{\textbf{Ok...still not sure how to prove this particular case}}}
%However, $\sigma_i V=q_0\sigma_i+q_1\mathbb{I}+q_2\sigma_i\sigma_j\notin span\{\mathbb{I}_2,\sigma_i,\sigma_j\}$

%%%%%%%%%%%%%%%%%%%%%%%%%%%%%%%%%%%%%%%%%%%%%%%%%%%%%%%%%%%%%%%%%%%%%%%%%%%%%%%%%%%%%%%%%%%%%%%%%%%%%%%%%%%%%%%%%%%%%%%%%%%%%%
\subsection{MUUB for n=2}
\noindent While all orthogonal (possibly distinct) bases  of $M(2,\mathbb{C})$ spans the same space, $M(2,\mathbb{C})$, the case for $n=2$ suggests the possibility of considering a number of possible orthogonal bases of which each basis could very well span a distinct subspace. Hence, constructing MUUB for one subspace naturally sets to question whether the same number of MUUB could be constructed in another subspace. 

Referring to Definition 2 and Proposition 1, we are however assured that,
%To this end, we define the equivalence between bases for distinct subspaces.\\  
%
%\noindent{\bf Definition 2}\\
%Consider two distinct orthogonal basis, $\mathcal{X}=\{X_1,X_2,...,X_n\}$ and $\mathcal{Y}=\{Y_1,Y_2,...,Y_n\}$ for $n\leq4$ spanning possibly distinct subspaces $M(\mathcal{X})$ and $M(\mathcal{Y})$ respectively. $\mathcal{X}$ is equivalent to $\mathcal{Y}$, $\mathcal{X}\equiv \mathcal{Y}$, provided that ${\red \zeta_j Y_j=UX_j}, \forall j=1,2,...,n$ for some unitary $U$ and ${\red |\zeta_j|=1}$.\\ 
%
%\noindent{\bf Proposition 3}\\
%Given the above definition for $\mathcal{X}\equiv \mathcal{Y}$ and $n\leq4$, $\mathcal{X}$ is {\red unbiased} with regards to a set $\mathcal{Z}=\{Z_1,Z_2,...,Z_n\}\subset M(\mathcal{X})$ if and only if $\mathcal{Y}$ is {\red unbiased} with regards to a set  $U\mathcal{Z}=\{UZ_1,UZ_2,...,UZ_n\}\subset M(\mathcal{Y})$.\\ 
%
%\noindent The above proposition is easily shown to be true by $|\text{Tr}(X_j^\dagger Z_k)|^2=|\text{Tr}((UX_j)^\dagger UZ_k)|^2=|\text{Tr}((\zeta Y_j)^\dagger UZ_k)|^2=|\text{Tr}((Y_j)^\dagger UZ_k)|^2$. 
%
%The case for $n=4$ has $M(\mathcal{X})=M(\mathcal{Y})$ and is trivially the case of changing basis in $M(2,\mathbb{C})$. 
%In the case of $n=2$, 
if we consider an arbitrary orthogonal basis defined by $\{A,B\}$ for a $2$ dimensional subspace of $M(2,\mathbb{C})$, it is equivalent to the basis $\mathcal{W}_0=\{\mathbb{I}_2,\omega\}$ where $\omega=A^\dagger B$ and spans a subspace, say, $M_1$ which would have the same number of MUUB for the subspace spanned by $\{A,B\}$.  
%nonequivalent orthogonal bases sets $\{\mathbb{I}_2,\sigma_1\}$, $\{\mathbb{I}_2,\sigma_2\}$ and $\{\mathbb{I}_2,\sigma_3\}$. 
 We then let $\omega$ be represented by a generic matrix like eq.(\ref{G}) with the factor $\alpha$ as $\gamma$. %=\{r_0\mathbb{I}_2+r_1\omega~|~r_0,r_1\in\mathbb{C}\}$. 

Let $W_0=r_0\mathbb{I}_2+r_1\omega\in M_1$  be a unitary matrix. With $\left|\text{Tr}(\mathbb{I}_2W_0)\right |^2=\left|\text{Tr}(\omega W_0)\right |^2=4|r_j|^2, \forall j$ and $W_0^\dagger W_0=\mathbb{I}_2$, we have 
$|r_j|^2=1/2$.
 Working with the same approach as we did for $n=4$, where we compare $W_0$ with a generic $2\times 2$ unitary matrix (say with a factor $\alpha=\beta$), we have 
\begin{eqnarray}
%r_0^2=1/2,\\
r_0^2=e^{2i\gamma}/2~,~r_1^2=e^{2i(\beta-\gamma)}/2.
\end{eqnarray} 
 The above, admits only $4$ sets of solutions for $(r_0,r_1)$ and given $(r_0,r_1)$ and $(-r_0,-r_1)$ provide for equivalent matrices, the number of MUUBs %to $(\mathbb{I}_2,\omega)$ are at most $2$. %Constructing a basis, $\mathcal{W}_1$, comprising of these 2 matrices becomes straightforward; 
% \begin{eqnarray}
% \mathcal{W}_1=\left\{\dfrac{\mathbb{I}_2+\omega}{\sqrt{2}},\dfrac{\mathbb{I}_2-\omega}{\sqrt{2}}\right \}.
 %{\red \mathcal{W}_1=\left\{\mathbb{I}_2+\omega/\sqrt{2},\mathbb{I}_2-\omega/\sqrt{2}\right \}.}
 %\end{eqnarray}
% We can thus conclude that the maximal number of {\red MUUB} 
for $n=2$ is 2.
Explicit examples %for $\omega$ 
in terms of Pauli matrices are $\{\mathbb{I}_2,\sigma_2\}$ which is unbiased with respect to $\{(\mathbb{I}_2\pm i\sigma_2)/\sqrt{2}\}$ and $\{\mathbb{I}_2,\sigma_3\}$ to $\{(\mathbb{I}_2\pm i\sigma_3)/\sqrt{2}\}$.

%%%%%%%%%%%%%%%%%%%%%%%%%%%%%%%%%%%%%%%%%%%%%%%%%%%%%%%%%
%%%%%%%%%%%%%%%%%%%%%%%%%%%%%%%%%%%%%%%%%%%%%%%%%%%%%%%%%

\section{MUUB and MUB of Maximally Entangled States}
\noindent  The study of unitary transformations, more generally quantum channels, is well known to be closely connected to that of maximally entangled states through the isomorphism between unitaries, $U$ on a $d$ dimensional Hilbert space, $\mathcal{H}_d$ and vectors, $|U\rangle\rangle$ (using the notation from \cite{sac,d1}) in $\mathcal{H}_d\otimes \mathcal{H}_d$,
\begin{eqnarray}
U\equiv \sum_{i}\sum_j \langle j|U|i\rangle |i\rangle|j\rangle=|U\rangle\rangle\in\mathcal{H}_d\otimes \mathcal{H}_d
\end{eqnarray} 
for some basis vectors $|i\rangle$,$|j\rangle$ of $\mathcal{H}_d$. With $|\langle\langle U_a|U_b\rangle\rangle|^2=\left|\text{Tr}(U_a^{\dagger}U_b)\right |^2$ and $|U\rangle\rangle/\sqrt{d}$ gives a maximally entangled state \cite{d2}, the search for MUUB is tantamount to looking for sets of maximally entangled states which not only form a basis for $\mathcal{H}_d\otimes\mathcal{H}_d$ but are also mutually unbiased to one another. Thus, refering to the previous sections, the dimensionality of $\mathcal{H}_2\otimes \mathcal{H}_2$ is $4$, i.e states are mutually unbiased if the absolute value of their inner product is given by $1/2$. Hence, with $|\mathbb{I}_2,\rangle\rangle/\sqrt{2}=|\Phi^+\rangle$, $|\sigma_1\rangle\rangle/\sqrt{2}=|\Psi^+\rangle$, $|\sigma_2\rangle\rangle/\sqrt{2}=|\Psi^-\rangle$ and $|\sigma_3\rangle\rangle/\sqrt{2}=|\Phi^-\rangle$, the maximal number of such MUBs are three and they are %\dfrac{1}{\sqrt{2}}|\mathbb{I}_2,\rangle\rangle,\dfrac{1}{\sqrt{2}}|\sigma_1\rangle\rangle,\dfrac{1}{\sqrt{2}}|\sigma_2\rangle\rangle,\dfrac{1}{\sqrt{2}}|\sigma_3\rangle\rangle 
\begin{eqnarray}\label{MES}
&&\mathfrak{B}_0=\left\{|\Phi^+\rangle,|\Psi^+\rangle,|\Psi^-\rangle,|\Phi^-\rangle\right\},\\\nonumber
\\\nonumber
&&\mathfrak{B}_k=\left\{\dfrac{|E_0\rangle+\sum_{j=1}^3c_j|E_j\rangle}{2}~|c_j\in\{\pm1\},\prod_{j=1}^3{c_j}=(-1)^k\right\},\\\nonumber
%\mathfrak{B}_2&=&\left\{\dfrac{|E_0\rangle+\sum_{j=1}^3d_j|E_j\rangle}{2}|d_j\in\{\pm1\},\prod_{j=1}^3{d_j}=1\right\}
\end{eqnarray}
for $k=1,2$ with $|E_0\rangle=|\Phi^+\rangle,|E_1\rangle=i|\Psi^+\rangle,|E_2\rangle=|\Psi^-\rangle,|E_3\rangle=i|\Phi^-\rangle$ (also known as `magic basis' \cite{mb}). It is worth noting that the requirement for entangled states as a basis would not allow for the construction of the maximal number of MUB for a four dimensional system (which would be 5). It was noted in \cite{MUB4}, that in constructing MUBs for two-qubit states, if one begins by constructing 3 Bell-type bases which are mutually unbiased, then one cannot construct two additional bases sets.

This connection can be extended to the case of MUUB for $n=2$ as the dimension of the subspace is $2$ where mutually unbiased states have the absolute value of their inner product as $1/\sqrt{2}$. The absolute value of the inner product between maximally entangled states, say $|U_a\rangle\rangle/\sqrt{2}$ and $|U_b\rangle\rangle/\sqrt{2}$ derived from differing MUUB in this case would result in $|\langle\langle U_b|U_a\rangle\rangle|/2=|\text{Tr}(U_b^\dagger U_a)|/2=1/\sqrt{2}$.% as $|\text{Tr}(U_b^\dagger U_a)|=\sqrt{2}$.  

%%%%%%%%%%%%%%%%%%%%%%%%%%%%%%%%%%%%%%%%%%%%%%%%%%%%%%%%%%%%%%%%%%%%%%%%%%%%%%%%%%%%%%%%%%%%%%%%%%%%%%%%%%%%%
\section{{ MUUB} for Prime numbered dimensional Quantum system}

It is natural to consider the construction of MUUBs on higher dimensional systems, $\mathcal{H}_d$; and much is hinted at from the qubit scenario. We begin with the issue of subspaces of $M(d,\mathbb{C})$ which may admit MUUBs (or rather those that do not). To this end, we provide the following theorem.
\\ 

\noindent{\bf Theorem 1}\\
Consider a subspace of $M(d,\mathbb{C})$ and let its dimensionality be $n$. If $n$ is neither $d$ nor $d^2$, then no MUUB can exist for such a subspace.\\ 

\noindent The proof for this follows from that of Section II. B. by having one basis consisting of the identity and some Pauli matrices and then demanding another which is MUUB to it to contain elements which is in the span of the former.
Let a basis $\mathcal{D}$ be defined (unitary operator basis \cite{gottes}) by
\begin{eqnarray}\label{XZ}
\{ X^{a_i}Z^{b_j} ~~|~~i\in[0,m],j\in[0,n]\}
\end{eqnarray}
for $m,n,a_i,b_i\in \mathbb{Z}_d$ and with $X$ and $Z$ as the generalized Pauli operators for $d$ dimensional system given by \cite{hall}. We fix $a_0=b_0=0$ so as $\mathcal{D}$ includes $\mathbb{I}_d$. We can represent this set by the set of pairs,
\begin{eqnarray}
C_\mathcal{D}=\{ (i,j)  ~~|~~i\in[0,m],j\in[0,n]\}
\end{eqnarray}
Consider an element $E_1$ from another basis $\mathcal{E}$ such that 
\begin{eqnarray}
E_1=\sum_i^m\sum_j^nq_{ij}X^{a_i}Z^{b_j} \in span (\mathcal{D})
\end{eqnarray}
for $q_{ij}\in\mathbb{C}$. Another element, say, $E_2\in \mathcal{E}$ can then be written as
\begin{eqnarray}
E_2=(\sum_i^m\sum_j^nq_{ij}X^{a_i}Z^{b_i})X^{a_k}Z^{b_l}\\\nonumber
=\sum_i^m\sum_j^nq_{ij}\eta_d X^{a_i+a_k}Z^{b_j+b_l}
\end{eqnarray}
for some element $X^{a_k}Z^{b_l}\in\mathcal{D}$ and $\eta_d=\exp{(2\pi i/d)}$. Requiring $\mathcal{E}$ be MUUB to $\mathcal{D}$, we require $E_2\in span (\mathcal{D})$ implying  $X^{a_i+a_k}Z^{b_j+b_l}\in\mathcal{D}$ or $({i+k},{j+l})\in C_\mathcal{D}$ for all $i,k\in[0,m]$ and $j,l\in[0,n]$. It's obvious to note that  $E_2\in span (\mathcal{D})$ only if $C_\mathcal{D}$ is closed under addition mod $d$ which gives $|\mathcal{D}|=d$ or $d^2$.

While this tells us about the nonexistence of MUUB in certain subspaces, it does not promise the existence of MUUBs for the remaining subspaces. 
To this extent, we only verify the existence of MUUB operating on $\mathcal{H}_d$ for the $d^2$ dimensional subspace with $d$ being a prime number  by noting the existence of MUB for maximally entangled states in $\mathcal{H}_d\otimes \mathcal{H}_d$. Given the recipe for constructing MUBs for maximally entangled states in the ref.\cite{yh}, we write a basis of maximally entangled states for $\mathcal{H}_d\otimes \mathcal{H}_d$ as $\{|U_d^{0,0}\rangle\rangle,...,|U_d^{d-1,d-1}\rangle\rangle\}$ with $|U_d^{n,m}\rangle\rangle$ given by
\begin{eqnarray}
|U_d^{n,m}\rangle\rangle=\dfrac{1}{\sqrt{d}}\sum_{p=0}^{d-1}\eta_d^{np}|(p+m)\bmod d\rangle |p'\rangle
\end{eqnarray}
for $n,m=0,1,...,d-1$, and $|(p+m)\bmod d\rangle$ and $|p'\rangle$ refer to states, each being elements of the orthonormal bases of their respective Hilbert spaces. A basis mutually unbiased with respect to the above could be one with elements given by $\mathbb{I}_d\otimes U_p|U_d^{n,m}\rangle\rangle$ with $U_p|p'\rangle\rightarrow|p''\rangle$
%\begin{eqnarray}
%|V_d^{n,m}\rangle\rangle=\dfrac{1}{\sqrt{d}}\sum_{p=0}^{d-1}\eta_d^{np}|(p+m)\bmod d\rangle |p''\rangle
%\end{eqnarray}
where $|\langle p'|p''\rangle|=1/\sqrt{d}$ (hence  $|\langle\langle U_d^{n,m}|U_p|U_d^{n,m}\rangle\rangle|=1/d$). As there would be only $d+1$ MUB for a $d$ dimensional system, we can conclude that this recipe provides for $d+1$ MUB for the maximally entangled states from $\mathcal{H}_d\otimes \mathcal{H}_d$ thus implying the same number for MUUBs acting on $\mathcal{H}_d$ for the $d^2$ dimensional subspace. This obviously provides for the minimal number of such MUUBs.  It was noted in \cite{scott} that the maximal number that such an MUUB can possibly have is $d^2-1$, though it remains unclear if this is saturated for all prime $d$.
Given all these, we have the following proposition.\\
\newline
\noindent{\bf Proposition 2}\\
The maximal and minimal number of MUUBs for a subspace of $M(d,\mathbb{C})$ with dimensionality $d^2$ is $d^2-1$ and $d+1$ respectively for a prime number $d$.\\ 
\newline
In the specific case of $d=2~(n=4)$, the minimal number of MUUB is in fact also the maximal, i.e. 3.

Constructing a maximal set of MUUB for any $d$ dimensional is obviously a challenge and following the construction for the qubit case, we consider the following. If a basis is given by Eq.(18) with m = n = d-1, and a unitary U is taken from a MUUB containing $X^{a_i} Z^{b_j}$, then, with $U^\dagger U=\mathbb{I}_d$ and $|\text{Tr}(U^\dagger X^{a_i}Z^{b_j})|^2=|\text{Tr}(U^\dagger X^{a_k}Z^{b_l})|^2, \forall{i,j,k,l}\in\mathbb{Z}_d$, we have 
\begin{eqnarray}
U=\sum q_{ij}X^{a_i}Z^{b_j}\Rightarrow |q_{ij}|=1/d^2
\end{eqnarray} 
and as $|\text{Tr}(U^\dagger X^{a_i}Z^{b_j})|^2=|q_{ij}\text{Tr}(\mathbb{I}_d)|^2$, we see that this reduces to the definition given for such untaries in \cite{scott}. It is now not difficult to consider a numerical search where one selects $q_{ij}=\exp{(2 \pi i/d)}^{t_{ij}}$ for some $t_{ij}\in\mathbb{Z}_d$ and search through all possible values for the possibilities of unitaries that fulfill the relevant criterion. Carrying this out for the case of $M(3,\mathbb{C})$ for example, we begin with a set of unitary matrices that span, $M(3,\mathbb{C})$, constructed from a set of unitary operator basis \cite{gottes} containing the identity $\mathbb{I}_3$ and the generalized Pauli matrices for qutrits,
\begin{eqnarray}\label{m3}
\mathcal{B}_{30}=\{X_3^jZ_3^k~~|~~j,k=0,1,2\}
\end{eqnarray} 
with $X_3$ and $Z_3$ the generalized Pauli operators for qutrits (explicitly given by \cite{hall}). % where the nonzero entries for $X$ are $X_{13}=X_{21}=X_{32}=1$ whereas for $Z$, they are $Z_{mm}=\eta^{2+m}$
We then find the following MUUBs,
\begin{eqnarray}\label{m34}
\mathcal{B}_{3l}=\{R_lX^jZ^k~~|~~j,k=0,1,2\}.
\end{eqnarray} 
 for $l=1,...,7$ with 
\begin{eqnarray}\label{m34}
%R_1=\dfrac{1}{3}[\mathbb{I}_3+\sum^A\sum_{j=0}^1(\eta^{jk} X^jZ^k)],\\\nonumber
R_1=\mathbb{I} + \eta_3 X + \eta_3^2  X^2 + \eta_3 Z + \eta_3^2 Z^2 \\\nonumber
+ \eta_3XZ + \eta_3^2 (XZ)^2 + \eta_3XZ^2 + \eta_3^2 (XZ^2)^2\\\nonumber
R_2= \mathbb{I} +X + \eta_3X^2 + Z + \eta_3 Z^2 \\\nonumber
+ \eta_3^2 XZ + (XZ)^2 + \eta_3^2 XZ^2  + (XZ^2)^2\\\nonumber
R_3= \mathbb{I} +X + \eta_3 X^2 + Z + \eta_3 ^2 Z^2 \\\nonumber
+XZ + \eta_3 ^2 (XZ)^2 +  \eta_3^2 XZ^2 + \eta_3^2   (XZ^2)^2  \\\nonumber
R_4=\mathbb{I} +X + \eta_3 X^2 + Z+ \eta_3^2 Z^2\\\nonumber
+ \eta_3 XZ + (XZ)^2 + \eta_3 XZ^2 +  \eta_3(XZ^2)^2\\\nonumber
R_5=\mathbb{I} +X + \eta_3 ^2 X^2 + Z + \eta_3 Z^2\\\nonumber
 + XZ+  \eta_3 ^2 (XZ)^2 +  \eta_3 XZ^2 +  (XZ^2)^2\\\nonumber
R_6=\mathbb{I} +  X + \eta_3 ^2 X^2 + Z + \eta_3 Z^2 \\\nonumber
+ \eta_3 XZ +  (XZ)^2 + XZ^2 + \eta_3 ^2 (XZ^2)^2 \\\nonumber
R_7=\mathbb{I} + X + \eta_3 ^2 X^2 + Z + \eta_3 ^2 Z^2 \\\nonumber
+ \eta_3 ^2 XZ + \eta_3 ^2 (XZ)^2 +   XZ^2 +  \eta_3  (XZ^2)^2
\end{eqnarray} 
%where $Q_1\in\{X,Z\}$, $Q_2\in\{XZ,XZ^2\}$ and $Q_3 =\mathcal{B}_{30}-\{\mathbb{I}_3,X,(XZ^2)^2\}$. $R_0$ is just $\mathbb{I}_3$. %Based on similar arguments of Section 2.B., we note that the only proper subspace (based on proper subset of $\mathcal{B}_{30}$) 
This saturates the maximal number of $8$ that has been showed to be achieved in \cite{scott}. As for the subspace of dimension $3$, we are able to construct 3 MUUBs. Beginning with say $\mathcal{D}_0=\{\mathbb{I}_3, A, A^2\}$, where $A\in\mathcal{B}_{30}-\{\mathbb{I}_3\}$, we can construct two sets of MUUB, $\mathcal{D}_1,\mathcal{D}_2$ given by 
\begin{eqnarray}
\mathcal{D}_j=\{D, DA, DA^2 \}
\end{eqnarray} 
where $D=\mathbb{I}_3+\eta_3^j A+\eta_3^j A^2$ for $j=1,2$.

%While making no claim to the maximal number of {\red MUUB} for the complete $M(d,\mathbb{C})$, based on the above, we propose the following conjecture:\\
%\newline
%\noindent {\bf Conjecture 1}\\
%{\red Given unitary operators acting on $H_d$, {\red MUUB}s can only be constructed for the entire vector space $M(d,\mathbb{C})$ and a $d$ dimensional subspace with the maximal number being $d^2-1$\cite{scott} and $d$ for the respective subspaces.} \\
%\noindent {\bf Conjecture 2}\\
%The maximal number of {\red MUUB}s for $M(d,\mathbb{C})$ is $d+1$ and for the $d$ dimensional subspace is $d$

%While making no claim to the maximal number of {\red MUUB} for the complete $M(d,\mathbb{C})$, we are tempted to think that the maximal is in fact $d+1$. This is at least verified in the case of $d=2$  for $n=4$ as well as suggested by the numerical search for the qutrit case. We also note another possible peculiarity, by extending the argument of Section 2.B, apart from the construction of {\red MUUB}s for $M(d,\mathbb{C})$, {\red MUUB}s can only exist for the subspace of dimensionality $d$.}  

%%%%%%%%%%%%%%%%%%%%%%%%%%%%%%%%%%%%%%%%%%%%%%%%%%%%%%%%%%%%%%%%%%%%%%%%%%%%%%%%%%%%%%%%%%%%%%%%%%%%%%%%%%%%%
\section{Applications}

Analogous to MUBs which sets constraints on state distinguishability, we consider in this section the issue of distinguishability of unitaries selected from a set of MUUBs on $\mathcal{H}_2$ and then proceed to consider its use in a QKD setup.

\subsection{Distinguishability of Unitaries}

%In this section we will study the distinguishability between the unitaries in the {\red MUUB}s; more precisely, this is equivalent to the estimation of an unknown unitary transformation chosen randomly from a complete set of {\red MUUB}. We will however limit our studies to the case of $n=4$. In doing so, we will make use extensively the work developed in \cite{chiri1,acin} for which the idea is simply as follows: consider having a black box that executes a unitary transformation, $U_m\in \cup_{i=0}^2 \mathcal{B}_i$, selected randomly from the union of the complete sets of {\red MUUB}s. One may submit a particular quantum state, $\rho$ (which may comprise of a qubit entangled to an ancillary system,),  through the black box and the resulting state, may be measured to provide information of the transformation. Given a single use of the box, the task requires not only an optimal input $\rho$ but also an optimal measurement of the output state. 

In distinguishing between unitaries selected randomly from the set $\text{SU}(d)$, Refs.\cite{chiri1, chiriT, acin} consider a black box executing the unitary transformation, where one may submit a particular quantum state (which may comprise of a qubit entangled to an ancillary system),  through the black box and the resulting state, may be measured to provide information of the transformation. Given a single use of the box, the task requires not only an optimal input but also an optimal measurement of the output state.  If one considers the set $\{U_g\}$ as a (projective) irreducible representation of a group $G$, a pure maximally entangled state, $|\Gamma\rangle$, may be used  as an optimal input \cite{d2} and the issue of discriminating between unitaries $U_g$ reduces to one of discriminating states in the group orbit \cite{chiriT}
\begin{eqnarray}
\{|\Gamma_g\rangle\langle\Gamma_g|=(U_g\otimes \mathbb{I}_R) |\Gamma\rangle\langle\Gamma | (U_g\otimes \mathbb{I}_R)^\dagger ~|~ g\in G\}.
\end{eqnarray}
It is necessary though to subscribe to the minimal discriminating requirement of Ref.\cite{chiriT} where $U_g=\lambda U_k\Rightarrow g=k$ for $\lambda\in \mathbb{C}, g,k\in G$. %Ref.\cite{acin} showed that the optimal input state can be chosen to be $\sum_i^d|i, i\rangle/\sqrt{d}\in\mathcal{H}_d\otimes\mathcal{H}_d$ being $\mathcal{H}_d$ a $d$ dimensional Hilbert space. 
%is indeed an irreducible representation of some group acting on $\mathcal{H}_2$ and
 
In considering the discrimination between elements of MUUB, we apply the same method above, though we restrict our study of to the case of $n=4$. More precisely, let us consider the scenario of selecting a unitary transformation randomly from the set of unitaries $\cup_{i=0}^2\mathcal{B}_i$  and we choose for the input state, $|\Psi\rangle=\sum_{i=0}^1|i, i\rangle/\sqrt{2}\in\mathcal{H}_2\otimes\mathcal{H}_2$.\footnote{ according to \cite{scott}, $\cup_{i=0}^2\mathcal{B}_i$ forms a 2-design.}
%\begin{eqnarray}
%|\Psi\rangle=\dfrac{1}{\sqrt{2}}|00\rangle+|11\rangle\in\mathcal{H}_2\otimes\mathcal{H}_2.
%\end{eqnarray}
It is easy to check that
\begin{eqnarray}
\forall~|\psi\rangle\in \mathfrak{B}_i, \forall U\in \mathcal{B}_j,%\\\nonumber 
~U\otimes I|\psi\rangle\in\mathfrak{B}_{(i+j)\bmod 2}
\end{eqnarray}
%where the operation $\oplus$ is addition modulo $2$. 
and the problem of discriminating between unitaries selected from the set $\cup_{i=0}^2\mathcal{B}_i$ reduces to an optimal discrimination of maximally entangled states forming mutually unbiased bases, i.e the group orbit here is $\{|\Gamma_g\rangle\langle\Gamma_g| ~|~|\Gamma_g\rangle\in\cup_{i=0}^2\mathfrak{B}_i\}$. %In this case, the estimation of unitaries from sets of {\red MUUB} coincides with the estimation of maximally entangled states from sets of MUBs.

In discriminating between the maximally entangled states above, we apply quite directly the method and use of extremal covariant measurements in \cite{chiriE} for quantum states of prime powered dimensions. Let us rewrite the states of $\mathfrak{B}_0$ with $|\Phi^+\rangle\equiv |0\rangle, |\Psi^+\rangle\equiv |1\rangle, |\Psi^-\rangle\equiv |2\rangle$ and $|\Phi^-\rangle\equiv |3\rangle$ with $\{|i\rangle,i\in \mathbb{F}_4\}$ as the `computational basis' in $\mathcal{H}_4$ and $\mathbb{F}_4$ is the finite field of cardinality $4$. Let us then consider the projective representation of the Abelian group $\bold{G}=\mathbb{F}_4\times\mathbb{F}_4$ as \cite{chiriE}
\begin{eqnarray}
R(\bold{G})=\{\mathcal{U}_q\mathcal{V}_w~|~(q,w)\in\mathbb{F}_4\times\mathbb{F}_4\}
\end{eqnarray}
where $\mathcal{U}_q|i\rangle=|i\oplus q\rangle~,~\mathcal{V}_w|i\rangle=\langle w,i\rangle|i\rangle$
%\begin{eqnarray}
%{\red \mathcal{U}_q|i\rangle=|i\oplus q\rangle~,~\mathcal{V}_w|i\rangle=\langle w,i\rangle|i\rangle}%\\\nonumber
%\mathcal{V}_w|i\rangle=\langle w,i\rangle|i\rangle
%\end{eqnarray}
with
$\langle w,i\rangle=\chi (w\odot i)$ where $\chi (x)$ is a nontrivial character of the additive group $\mathbb{F}_4$. The operations $\oplus$ and $\odot$ are of course field addition and multiplication respectively. Writing $x$ as a tuple $(s_1,s_2), s_i\in\{0,1\}$, we choose $\chi (x)$ as $\exp{(\pi i s_1)}$ \cite{kr}. 
Considering the state (call it inital), $|0\rangle\langle 0|$, it can be easily checked that its orbit under $R(\bold{G})$ is indeed $\{|i\rangle\langle i|,i\in \mathbb{F}_4\}$. This holds similarly for the states from $\mathfrak{B}_1$ and $\mathfrak{B}_2$ where the orbits under $R(\bold{G})$ for any one `intial' state in one basis is the set of all states from the respective bases and the stability groups for the initial states are nontrivial. From \cite{chiriE}, with $R(\bold{G})$ being irreducible, an extremal positive operator valued measure is the group orbit of a single operator and we may conclude that, with the choice of basis, $\mathfrak{B}_0,\mathfrak{B}_1,\mathfrak{B}_2$ being equally probable, the optimal measurement operator is the orthogonal measurement onto any one of the bases. 
The average estimation fidelity \cite{acin} in discriminating these states, can then be written as
\begin{eqnarray}
\frac{1}{12}\sum_{u,s} P(|s\rangle ||u\rangle)|\langle s| u\rangle|^2
\end{eqnarray}
where $P(|s\rangle ||u\rangle)$ is the probability of a measurement of state $|u\rangle$ resulting in $|s\rangle$. It is easy to show that the average fidelity of discriminating between states selected randomly from $\cup_{i=0}^2\mathfrak{B}_i$ is really $1/2$. This is in fact the same maximal value for the average estimation fidelity of a completely unknown maximally entangled states in $\mathcal{H}_2\otimes\mathcal{H}_2$ immediately calculated from \cite{acin} as well as that for a completely unknown unitary in $\text{SU}(2)$ as from \cite{bisio}.

%%%%%%%%%%%%%%%%%%%%%%%%%%%%%%%%%%%%%%%%%%%%%%%%%%%%%%%%%
%%%%%%%%%%%%%%%%%%%%%%%%%%%%%%%%%%%%%%%%%%%%%%%%%%%%%%%%%
\subsection{MUUB in a Quantum Key Distribution Setup}
The idea of `bases' for unitary transformation was mentioned in \cite{bisio} and later also used in \cite{jss} to highlight richer points as opposed to a prepare and measure like QKD scheme. From the above sections describing the `unbiasedness' of such bases in a more precise way, applying it to two-way QKDs is the most natural step in generalizing such protocols to include all MUUBs. %This further highlights the essence of such a protocol. 
The basic structure of the protocol remains; Bob would submit a qubit prepared in a particular basis to Alice who would encode on it with a unitary transformation before sending the qubit back to Bob for measurements. %Ref.\cite{jss} exemplifies the application of {\red MUUB} for the 2-dimensional subspace where only $2$ {\red MUUB}s exist. The protocol proposed in \cite{chiri2} exhibits the use of 2 out of 3 {\red MUUB}s for the case $n=4$. 

For the sake of clarity, we propose a protocol which is a generalization (in terms of the encoding) of one proposed and analyzed in \cite{norm}. For encoding purposes, Alice would randomly select an operator from her `bases' $\mathcal{B}_0,\mathcal{B}_1$ and $\mathcal{B}_2$. Binary values are assigned to each operator in $\mathcal{B}_0$ as in \cite{norm}, i.e. $0,1,1,0$ to $\mathbb{I}_2,\sigma_1,\sigma_2,\sigma_3$ respectively, which can be done as well for $\mathcal{B}_1$ and $\mathcal{B}_2$. Bob would submit a qubit prepared in one of two MUBs and measures the returned qubit in a randomly selected basis; either the same as the one he prepared in or one mutually unbiased to it. While $\mathcal{B}_0$ would retain the bases of Bob's qubit states, $\mathcal{B}_1,\mathcal{B}_2$ would shift them to a basis mutually unbiased. At the end of the protocol, Alice would announce (on a public channel) which of the the `bases' she had used for a given round of the protocol. Alice would further disclose if she had used either one of two subsets from her bases choice where each subset would contain elements that can be distinguished perfectly by Bob's measurements (subject to Bob measuring the qubit in the correct basis). As a quick example from \cite{norm}, the subsets $\{\mathbb{I}_2,\sigma_2\}$ and $\{\sigma_1,\sigma_3\}$. Hence, $1/3$ of the time Bob would retain his measurements where he can determine Alice's encoding conclusively for key purposes while discarding the rest (instances where he measures a qubit in a wrong basis). 

To provide an insight into the security of such a protocol, we consider the simplest strategy for an eavesdropper, Eve, to ascertain Alice's encoding. She would hijack Bob's qubit en route, submit a Bell state to Alice instead to estimate the unitary used and then apply her estimation on Bob's qubit before returning to him. Subsequent to Alice's disclosure, Eve's gain would only be $1/3$ while inducing an equal amount of error. While obviously a more involved strategy should be considered, for an error less than $1/2$ between Alice and Bob, Eve's gain would never achieve unity due to the inability to distinguish between MUUBs perfectly for a single use. A proper security analysis is however, beyond the scope of this work.

%%%%%%%%%%%%%%%%%%%%%%%%%%%%%%%%%%%%%%%%%%%%%%%%%%%%%%%%%
%%%%%%%%%%%%%%%%%%%%%%%%%%%%%%%%%%%%%%%%%%%%%%%%%%%%%%%%%

%%%%%%%%%%%%%%%%%%%%%%%%%%%%%%%%%%%%%%%%%%%%%%%%%%%%%%%%%
%%%%%%%%%%%%%%%%%%%%%%%%%%%%%%%%%%%%%%%%%%%%%%%%%%%%%%%%%

\section{Conclusion}
 
Generalizing the notion of MUUB formalized in \cite{scott}, we provide a definition for sets of unitary transformations forming bases for subspaces of $M(d,\mathbb{C})$, such that the elements in one basis are `mutually unbiased' with respect to elements in another. %We have formalized in this work, the notion of unitary transformations forming bases of $M(d,\mathbb{C})$, for which the elements in one basis are `mutually unbiased' with respect to elements in another. 
The essence of the definition is in capturing the notion of distinguishability between unitary transformations based on their actions on quantum states. We explicitly construct such bases for the qubit case and show how such a construction gives the maximal number of such bases as $3$ for the 4-dimensional vector space of $M(2,\mathbb{C})$ and $2$ for a 2-dimensional subspace of $M(2,\mathbb{C})$. Subsequent to that, we consider the case for unitary operators acting on prime numbered dimensional systems and prove the nonexistence of MUUB for subspaces of $M(d,\mathbb{C})$ of dimensionality other than $d$ or $d^2$. Subscribing to a simple numerical search, we construct MUUBs for qutrits in all possible subspaces. 
 
In a bid to see MUUBs in action beyond its construction, we note that in the case for qubits, estimating a unitary selected randomly from the full set of MUUBs is equivalent to the estimation of a maximally entangled states selected randomly from the maximal number of MUBs with the average estimation fidelity as $1/2$; i.e. equal to the case for estimating a completely unknown maximally entangled states or a completely unknown unitary from $\text{SU}(2)$. We then consider this in a QKD setup. These in fact would be beyond the role that MUUBs have shown to play in unitary 2-designs \cite{scott}.
  
There are obviously various other interesting directions this work may be extended to, including an infomation-disturbance tradeoff in the estimation of such unitarmaximally entangled statesies (or its isomorphic equivalent of maximally entangled states), a more specific scenario of \cite{bisio} (\cite{MFS}) as well as a proper understanding of possible entropic bounds in such estimations. Immediate applications of such studies include a possibly more thorough study of quantum cryptography as described in Section V. More immediate issues would include a deeper understanding of MUUB acting on higher dimensional quantum systems.

%%%%%%%%%%%%%%%%%%%%%%%%%%%%%%%%%%%%%%%%%%%%%%%%%%%%%%%%%%%%%%%%%%%%%%%%%%%%%%%%%%%%%%%%%%%%%%%%%%%%%%%%%%%%%
\section*{Acknowledgement}
One of the authors, J. S. S would like to thank H. Zainuddin and S. Karumanchi for fruitful discussions. He is grateful to the University of Camerino for the kind hospitality during the time this work was conceived and  for financial support by Ministry of Higher Education (Malaysia), Fundamental Research Grant Scheme FRGS14-152-0393. as well as the University's Research Management Centre for their assistance.

\end{document}